\documentclass[aps,twocolumn,showpacs,amsmath,amssymb]{revtex4} %PRL!!!

\usepackage[latin1]{inputenc} %%SUOMI
\usepackage{amsthm}
\usepackage{latexsym}
\usepackage{amsfonts}
\usepackage{bbm,bm,dsfont}
\usepackage{color}

\newcommand{\R}{\mathbb R}

\newcommand{\C}{\mathbb C}
\newcommand{\Q}{\mathsf Q}

\newcommand{\hi}{\mathcal{H}} %Hilbert space
\newcommand{\tr}[1]{\mathrm{tr}\left[#1\right]} %trace
\newcommand{\ket}[1]{|#1\rangle} %ket
\newcommand{\kb}[2]{|#1\,\rangle\langle\,#2|} %ketbra
\newcommand{\Mo}{\mathsf{M}} %generic observable
\newcommand{\No}{\mathsf{N}} %generic observable
\newcommand{\Po}{\mathsf{P}} %generic observable
\newcommand{\A}{\mathsf{A}}

\def\<{\langle}
\def\>{\rangle}
\def\d{{\mathrm d}}
\newcommand{\fii}{\varphi}

\begin{document} 

\title[Complete quantum measurements]{Complete quantum measurements break entanglement}

%\title[Complete quantum measurements]{Complete quantum measurements break entanglement}

%

\author{Juha-Pekka Pellonp\"a\"a}
\email{juhpello@utu.fi}
\address{Turku Centre for Quantum Physics, Department of Physics and Astronomy, University of Turku, FI-20014 Turku, Finland}

\begin{abstract}
Complete measurement of a quantum observable (POVM) is a measurement of the maximally refined version of the POVM.
Complete measurements give information on multiplicities of measurement outcomes and can be viewed as state preparation procedures. Moreover,
any observable can be measured completely. In this Letter, we show that a complete measurement breaks entanglement completely between the system, ancilla and their environment. 
Finally, consequences to the quantum Zeno effect and
complete position measurements are discussed.
\end{abstract}


\maketitle

From the perspective of a single observer, a quantum mechanical system can be divided into three parts: the system $S$ under observation, an ancillary or probe system $A$ describing the measurement apparatus, and their environment $E$ (with Hilbert spaces $\hi_S$, $\hi_A$, and $\hi_E$, respectively).
In the standard quantum measurement theory, the influence of $E$ is usually neglected by assuming that i) $E$ does not interact with $S+A$ during the measurement of an observable of $S$ and ii) the initial state (density matrix) $\omega$ of the total system $E+S+A$ (with $\hi_E\otimes\hi_S\otimes\hi_A$) is factorized (i.e.\ of the form $\omega=\omega_E\otimes\omega_S\otimes\omega_A$).
Then the measured system $S$ interacts only with $A$, resulting in an entangled state of $\hi_S\otimes\hi_A$
\footnote{An entangled state of a composite system 
is not separable, i.e.\ it cannot be written as a convex combination (mixture) of factorized (product) states.}.
 Since the state before measurement is assumed to be factorized, the entanglement occurs only in the bipartite subsystem $S+A$ and
$E$ can be isolated from $S+A$. It follows from von Neumann's projection postulate that the states of $S+A$ and $E+S+A$ are not entangled immediately after the measurement.

The condition of locality i) is quite reasonable, e.g., if the duration of the measurement is short compared to the interaction time between $E$ and $S+A$ and if the measurement interaction is strong enough to dominate the dynamics of the total system \footnote{In addition, if the system interacts with its environment during a measurement then one can redefine the Hilbert space $\hi_S$ of the system in such a way that it contains the interacting part of the environment.
}, but ii) is not so clear. 
Namely, ii) requires that the preparation procedure of the initial state is entanglement breaking. However, in practice, the preparation may be incomplete, $S+A$ can interact with $E$ between the preparation and measurement, or the measurement can be a part of the chain of measurements so that the separability of the initial state is not guaranteed.

In this Letter, we give up assumption ii) and consider measurements (and preparations) which break entanglement.
We will see that the so-called {\it complete measurements} \cite{PeCOMP} break entanglement and can be viewed as entanglement breaking state preparation procedures.

Any quantum observable can be measured completely \cite{PeCOMP}.
For example, suppose that we want to measure the energy (i.e.\ Hamiltonian operator $H$) of a quantum system $S$. The spectrum of $H$ may be degenerate so that then the measurements of $H$ cannot be seen complete since they do not give `information' about degeneracies of energy states. Now the complete measurement  of energy would be a measurement of the maximally refined version of $H$ since its outcome space contains also degeneracies of $H$. One can measure the maximally refined $H$ by first measuring $H$ and then some other `multiplicity' observable. Such a complete measurement can be seen as a preparation of a new measurement even if the specific form of the measurement interaction is not known. 
If the measurement of $H$ is not complete and the 
reduced state of the system is an eigenstate of $H$ the state of $E+S$ can still be entangled after the measurement.
We discuss this further at the end of this Letter.
Next we briefly recall the mathematical description of a quantum observable as a \emph{(normalized) positive operator valued measure} (POVM) 
\cite{Da,BuLaMi,TeZi,Ho} and study entanglement breaking measurements.

Consider a quantum system $S$ with a Hilbert space $\hi_S$ and suppose that the measurement outcomes form a set $\Omega=\{x_1,x_2,\ldots\}$ \footnote{Note that $\hi_S$ can be infinite dimensional and the number $\#\Omega$ of the elements of $\Omega$ can be infinite everywhere in this Letter.}.
A POVM $\Mo$ is a collection of positive operators $\Mo_i$ such that
$\sum_{i=1}^{\#\Omega}\Mo_i=I_S$ (the identity operator of $\hi_S$).
Now for every state $\varrho$ of $S$, the mapping
$x_i\mapsto p^i_\varrho=\tr{\varrho\Mo_i}$
is a probability distribution and $p^i_\varrho$ is the probability of getting a measurement outcome $x_i$ when the system is in the state $\varrho$ and a measurement of $\Mo$ is performed. 
A POVM $\Mo$ is {\it rank-1} if any $\Mo_i$ can be written in the form $\Mo_i=\kb{d_i}{d_i}$ where $d_i\in\hi_S$, and $\Mo$ is called a \emph{projection valued measure} (PVM) if $\Mo_i^2\equiv\Mo_i$.
Recall that (real) PVMs or spectral measures correspond to self-adjoint operators.

A (minimal) {\it measurement model} \cite{BuLaMi,TeZi,Oz84}
for a rank-1 POVM $\Mo$ is given by a unitary measurement interaction $U_{SA}$ on $\hi_S\otimes\hi_A$, 
\begin{equation*}
U_{SA}(\psi\otimes\xi)=\sum_{i=1}^{\#\Omega}\<d_{i}|\psi\>\fii_{i}\otimes e_i,
\qquad\psi\in\hi,
\end{equation*}
where the unit vectors $\fii_i\in\hi_S$ and $\xi\in\hi_A$ are fixed and the vectors $e_i$ form an orthonormal basis of the ancillary Hilbert space $\hi_A$ of dimension $\#\Omega$ \cite{PeCOMP,Pe12}.
The pointer PVM $\Po$ is $\Po_i=\kb{e_i}{e_i}$ so that
if the system (resp.\ apparatus) is in the state $\varrho$ (resp.\ $\xi$) before the measurement and the compound `object-apparatus' ($S+A$) system is in the factorized state $\varrho\otimes\kb{\xi}{\xi}$ then the (non-normalized) state after the measurement is
\begin{eqnarray*}
%\omega^i_\varrho&=&
&&
(I_S\otimes P_i)U_{SA}(\varrho\otimes\kb{\xi}{\xi})U_{SA}^*(I_S\otimes P_i) \\
&&\qquad=\;p^i_\varrho\kb{\fii_i}{\fii_i}\otimes\kb{e_i}{e_i}
\end{eqnarray*} 
when $x_i$ is obtained (the projection postulate), i.e.\ the ancilla has collapsed into the `eigenstate' $e_i$ whereas the system is in the (posterior) state $\fii_i$.
Hence, in the measurement process,  the non-entangled state $\varrho\otimes\kb{\xi}{\xi}$ is transformed into the entangled state $U_{SA}(\varrho\otimes\kb{\xi}{\xi})U_{SA}^*$ and finally the reading of $x_i$ breaks entanglement, the final state 
$\kb{\fii_i}{\fii_i}\otimes\kb{e_i}{e_i}$ being factorized.

From the perspective of the system, this measurement process can be described in terms of a (rank-1) {\it instrument} \cite{Da,BuLaMi,TeZi,DaLe}
$$
\mathcal I_i(\varrho)=\kb{\fii_i}{d_i}\varrho\kb{d_i}{\fii_i}=p^i_\varrho\kb{\fii_i}{\fii_i}
$$
which gives both the posterior states $\fii_i$ of the system and the measurement outcome probabilities $p_\varrho^i=\tr{\mathcal I_i(\varrho)}$.
Note that $\Mo$ is a PVM exactly when $\<d_i|d_j\>=\delta_{ij}$ and vectors $d_i$
form a basis of $\hi_S$ \cite{Pe11,HyPeYl}. Then, by choosing $\fii_i=d_i$, the instrument $\cal I$ is the
von Neumann-L\"uders \cite{BuLaMi} instrument, ${\cal I}_i(\varrho)=\Mo_i\varrho\Mo_i$, and
the quantum channel $\mathcal I_\Omega(\varrho)=\sum_{i}\mathcal I_i(\varrho)=\sum_i\kb{d_i}{d_i}\varrho\kb{d_i}{d_i}$ describes pure decoherence \cite{TeZi}
in the decoherence basis $\{d_i\}$.

Suppose then that the composite system $S+A$ has an environment $E$ with a Hilbert space $\hi_E$, and the environment does not interact with $S+A$ during the measurement. The Hilbert space of the total system is $\hi_E\otimes\hi_S\otimes\hi_A$ and the measurement interaction is $I_E\otimes U_{SA}$. 
It is interesting to see what happens if $E+S$ is in some (entangled) state $\omega_{ES}$ (of $\hi_E\otimes\hi_S$) before the measurement. If $A$ is in the state $\xi$ then the (non-normalized) state after the measurement is
\begin{eqnarray*}
&&(I_E\otimes I_S\otimes P_i)(I_E\otimes U_{SA})(\omega_{ES}\otimes\kb{\xi}{\xi})\times\\
&&\;\qquad\times\,(I_E\otimes U_{SA})^*(I_E\otimes I_S\otimes P_i) \\
&&\quad=\;({\rm id}\otimes\mathcal I_i)(\omega_{ES})\otimes\kb{e_i}{e_i} \\
&&\quad=\;{\rm tr}_{\hi_S}\Big[\big(I_E\otimes\sqrt{\Mo_i}\big)\omega_{ES}\big(I_E\otimes\sqrt{\Mo_i}\big)\Big]
\otimes\\
&&\;\qquad\otimes\,
\kb{\fii_i}{\fii_i}\otimes\kb{e_i}{e_i} 
\end{eqnarray*} 
if $x_i$ is registered, so that the {\it state is factorized (non entangled).} 
But this means that the operation $\mathcal I_i$ is {\it entanglement breaking}
\cite{HoShRu}.

On the first hand, it can be shown \cite{HeWo,Pe12} that any instrument related to any measurement of a rank-1 POVM $\Mo$ is
\begin{equation}
\label{instru}
\mathcal I_i(\varrho)=\tr{\varrho\Mo_i}\sigma_i
\end{equation}
where the states $\sigma_i$ are fixed, that is,
 $\mathcal I_i$ is entanglement breaking \cite{HoShRu}.
Also this means that the instrument $\cal I$ describes a {complete} measurement of $\Mo$ in the sense that the output states are completely known whatever the input state $\varrho$ is: if $x_i$ is registered then the state $\sigma_i$ is obtained \cite{PeCOMP}.

On the other hand, if all instruments related to a POVM $\Mo$ are of the form \eqref{instru} then $\Mo$ is rank-1 \cite{HeWo}.
Hence, we may conclude that
\begin{quote}
a POVM admits only entanglement breaking measurements if and only if it is rank-1.
\end{quote}
Indeed, this holds also for arbitrary (i.e.\ nondiscrete) rank-1 POVMs.
If $\Omega$ is arbitrary (e.g.\ $\R$) and $\Mo$ is a rank-1 POVM with the outcome space $\Omega$, then all its instruments are of the form
\begin{equation}\label{instru11}
\mathcal I_X(\varrho)=\int_X \sigma(x)\,\tr{\varrho\Mo(\d x)}
\end{equation}
where $\sigma(x)$ are states and $X\subseteq\Omega$ \cite{Pe12}.
Following Holevo {\it et al}.\ \cite{HoShWe,Ho08} one sees that the above operations $\mathcal I_X$ are entanglement breaking.
Recall that $\tr{\mathcal I_X(\varrho)}=p_\varrho(X)=\tr{\varrho\Mo(X)}$ is the probability of getting outcome $x$ which belong to $X$ and
$$
\varrho_X=\mathcal I_X(\varrho)/\tr{\mathcal I_X(\varrho)}
=p_\varrho(X)^{-1}\int_X  \sigma(x) \d p_\varrho(x)
$$
is the averaged state after a sequence of measurements which can be interpreted as follows:
if one gets an outcome $x$ then $\sigma(x)$ is the output state and $\varrho_X$ is a statistical mixture of output states $\sigma(x)$ when outcomes lie in $X$ \cite{PeCOMP}.
It is easy to show \cite{HoShWe} that $({\rm id}\otimes\mathcal I_X)(\omega_{ES})$ is separable (non entangled) whatever the initial state $\omega_{ES}$ of $E+S$ and $X$ are.

In addition, $\mathcal I$ (or any measurement of a rank-1 $\Mo$) can be viewed as a state preparator: $\varrho$-independent states $\sigma(x)$ (or their mixtures $\varrho_X$) are prepared with probabilities given by $p_\varrho$. In practice, one chooses $X$ as `small' as possible so that $\varrho_X\approx\sigma(x)$ when $x\in X$, and selects only output states which correspond to outcomes $x\in X$.
Moreover, the quantum channel $\mathcal I_\Omega$ of the measurement 
\footnote{Note that instruments of the form \eqref{instru} or \eqref{instru11} are called {\it nuclear} \cite{Oz85} and their channels $\mathcal I_\Omega$ {\it separable} (or they are of the Holevo form). A separable channel is an example of local operation and classical communication (LOCC). }
can be simulated by a classical channel: The sender performs a measurement on the input state $\varrho$ and sends the outcome $x$ via a classical channel to the receiver who then prepares $\sigma(x)$ \cite{HoShRu,TeZi}.
 To conlude,
\begin{quote}
a POVM admits only `measure-and-prepare' measurements if and only if it is rank-1.
\end{quote}
The next question is how rank-1 POVMs can be obtained from arbitrary POVMs.

Indeed, for any POVM $\Mo$, one can measure its maximally refined rank-1 version $\Mo^1$ by performing a measurement of $\Mo$ followed by the von Neumann-L\"uders measurement of a discrete self-adjoint operator \cite{PeCOMP}:
For simplicity we consider discrete POVMs.
Now any POVM $\Mo$ can be written in the form
$$
\Mo_i=\Mo\big(\{x_i\}\big)=\sum_{k=1}^{m_i}|d_{ik}\>\<d_{ik}|
$$
where $m_i$ is the {\it rank} of the effect $\Mo_i$ or the {\it multiplicity} of the outcome $x_i$ (and vectors $\{d_{ik}\}_{k=1}^{m_i}$ are linearly independent) \cite{PeCOMP,Pe11,HyPeYl}.
Immediately one sees that $\Mo$ can be {\it maximally refined} into rank-1 POVM $\Mo^1$ whose value space $\Omega_\Mo$ consists of pairs $(x_i,k)$ and 
\begin{equation}\label{kljsdhklfh}
\Mo^1_{ik}=\Mo^1\big(\{(x_i,k)\}\big)=|d_{ik}\>\<d_{ik}|.
\end{equation}
Hence, a measurement of $\Mo^1$ contains also information about multiplicities $k$ of measurement outcomes $x_i$ of $\Mo$.

Let $K$ be the maximal rank of $\Mo$, i.e.\ $m_i\le K$ for all $i$, and
pick a self-adjoint operator $N$ of $\hi_S$ with at least $K$ eigenvalues $a_k\in\R$ and let $\phi_k$ be some (orthonormal) eigenvectors, i.e.\ $N\phi_k=a_k\phi_k$. 
As before, define a measurement interaction
$$
U_{SA}(\psi\otimes\xi)=\sum_{i=1}^{\#\Omega}\sum_{k=1}^{m_i}\<d_{ik}|\psi\>\phi_{k}\otimes e_i%=\sum_{i=1}^{\#\Omega}\A_i\psi\otimes e_i
$$
giving an instrument
$$
{\mathcal I}_i(\varrho)=\sum_{k,l=1}^{m_i}\<d_{ik}|\varrho|d_{il}\>
|\phi_{k}\>\<\phi_{l}|
$$
for which
$$
\tr{{\mathcal I}_i(\varrho)}=\sum_{k=1}^{m_i}\<d_{ik}|\varrho|d_{ik}\>=\tr{\varrho\Mo_i}=p^i_\varrho.
$$
Immediately after performing this measurement of $\Mo$, measure $N$ by using the von Neumann-L\"uders measurement \cite{BuLaMi} for $N$ to get
$$
{\mathcal I}_{ik}(\varrho)=\mathsf N_k{\mathcal I}_i(\varrho)\mathsf N_k=\<d_{ik}|\varrho|d_{ik}\>|\phi_{k}\>\<\phi_{k}|,
$$
an instrument implementing $\Mo^1$. (Here $\mathsf N_k$ is the eigenprojection of $N$ corresponding to $a_k$.)
Thus, one gets $x_i$ in the first measurement of $\Mo$ and $k$ in the second measurement of $N$ with the probability $$p^{ik}_\varrho=\tr{\varrho\Mo_{ik}^1}=\<d_{ik}|\varrho|d_{ik}\>=\tr{{\mathcal I}_{ik}(\varrho)}$$
and the post measurement state is $|\phi_{k}\>\<\phi_{k}|$, i.e.\
the state of the system has collapsed into the eigenstate $\phi_k$ of $N$
and the measurement is completed.
Similarly, it can be shown \cite{PeCOMP} that one can measure the maximally refined (rank-1) version of an arbitrary POVM leading to a conclusion:
\begin{quote}
Any quantum observable $\Mo$ can be measured {\it completely,} i.e.\ one can measure the maximally refined version $\Mo^1$ of $\Mo$. Complete measurements 
give complete information on multiplicities of measurement outcomes, can be viewed as state preparations thus completing measurement chains, and completely break entanglement.
\end{quote}
Note that $\Mo$ above is arbitrary and $N$ is quite arbitrary so that they do not have to form a complementary pair (like position and momentum) or commute.
It is a standard result that any two mutually commuting observables have a joint observable and can be measured together but arbitrary observables can always be measured sequentially. 
%It is a little surprising that only one extra observable is needed to measure $\Mo^1$.

As an example, we consider a spin-$\frac{1}{2}$ particle moving on a line $\Omega=\R$.
The Hilbert space $\hi_S$ of the particle consist of wave functions
$$
\psi(x)={{\psi_+(x)}\choose{\psi_-(x)}}.%=\Psi_+(x)+\Psi_-(x)
$$
%$$\Psi_+(x)=\psi_+(x){1\choose0},\qquad\Psi_-(x)=\psi_-(x){0\choose1}$$
Define generalized vectors $d_{\pm}(x)$ by
$\<d_\pm(x)|\psi\>=\psi_\pm(x)$ \cite{HyPeYl} and projections $\No_{\pm}\psi=\psi_\pm$.
The position observable (PVM) of the particle is then
$$
\Q(X)=\int_X \big[|d_+(x)\>\<d_+(x)|+|d_-(x)\>\<d_-(x)|\big]\d x
$$
where $X\subseteq\R$, and the spin operator is
$$
N=\frac{\hbar}{2}\left(\begin{array}{cc}1 & 0 \\0 & -1\end{array}\right)
=\frac{\hbar}{2}\No_+-\frac{\hbar}{2}\No_-.
$$
The measurement models of $\Q$ are determined in \cite{Pe12} so that  next we consider a special case.

Let $\fii_0:\,\R\to\C$ be a (square integrable) function and
$$
\psi_0={{\fii_0}\choose{\fii_0}}=\ket++\ket-
$$
where
$$
\ket+=\fii_0{1\choose0}={\fii_0\choose0},\qquad
\ket-=\fii_0{0\choose1}={0\choose \fii_0}.
$$
For example, we may choose $\fii_0$ to be the `single mode vacuum' $\ket0=h_0$,
$$
h_0(x)=\frac1{\sqrt[4]\pi}\,e^{-x^2/2},
$$
so that 
$
\psi_0={{\ket0}\choose{\ket0}}
$
is the `two mode vacuum'.
A minimal position measurement \footnote{
Now $\hi_A=L^2(\R)$ (the wave functions $\R\to\C$), the pointer PVM $\Po$ is the usual position observable, $\Po(X)=\chi_X$, and the interaction is $[U(\psi\otimes\xi)](x)=\ket+\otimes\psi_+(x)+\ket-\otimes\psi_-(x)$ where $\xi$ is the initial state of the ancilla $A$.} yields an instrument
$$
\mathcal I_X(\varrho)=\int_X \A(x)\varrho\A(x)^*\d x
$$
where
$\A(x)=\kb{+}{d_+(x)}+\kb{-}{d_-(x)}$, i.e.\
$$
\A(x)\psi=\psi_+(x)\ket+ +\psi_-(x)\ket-.
$$
Now $\tr{\mathcal I_X(\varrho)}=\tr{\varrho\Q(X)}$.
%$\A(x)\ket\pm=h_0(x)\ket\pm.$

Suppose that the Hilbert space of the environment $E$ is
$\hi_E=\hi_S$ and the initial state of the composite system $E+S$ is an entangled
Bell state
$$
\Psi=\frac{1}{\sqrt2}\big(\ket{++}+\ket{--}\big).
$$
After the measurement of $\Q$ the (non-normalized) state is
$$
({\rm id}\otimes\mathcal I_X)(\kb{\Psi}{\Psi})=\int_X|\fii_0(x)|^2\d x\;\kb{\Psi}{\Psi}
$$
showing that the post measurement state is the Bell state $\Psi$ (for all $X$). Hence, the position measurement is not entanglement breaking.

One can complete the position measurement by measuring also the spin of the particle: for example, if one gets $\hbar/2$ in the spin measurement the state collapses to
$$
(I_E\otimes\mathsf N_+)\kb{\Psi}{\Psi}(I_E\otimes\mathsf N_+)=\frac{1}{2}\kb{++}{++}
$$
and the sequential measurement of $\Q$ and $N$ is entanglement breaking.
The instrument is now
$$
\mathcal I_{X,\pm}(\varrho)=\mathsf N_\pm\mathcal I_{X}(\varrho)\mathsf N_\pm
=\int_X\<d_\pm(x)|\varrho|d_\pm(x)\>\d x\kb{\pm}{\pm}
$$
so that the measurement outcome probability
\begin{eqnarray*}
\tr{\mathcal I_{X,\pm}(\varrho)}&=&\int_X\<d_\pm(x)|\varrho|d_\pm(x)\>\d x \\
&=&\tr{\varrho\,\Q^1(X\times\{\pm\hbar/2\})}
\end{eqnarray*}
where 
$$
\Q^1(X\times\{\pm\hbar/2\})=\int_X |d_\pm(x)\>\<d_\pm(x)|\,\d x
$$
is the maximally refined position observable of the particle \footnote{This construction can easily be extended to the general case where a (nonrelativistic) particle of spin $j$ moves on an arbitrary space (Riemannian manifold) $\Omega$ with a metric $g_{\mu\nu}$.
Now $\hi_S$ consists of $\C^{2j+1}$-valued wave functions on $\Omega$ which are square integrable with respect to the volume form generated by $g_{\mu\nu}$.
Hence, the {\it position measurements of a particle are never complete} (if $j\ne 0$) but, when combined with spin measurements, they form complete (rank-1) observables.}.

Obviously, this example can be applied to quantum optics where $\hi_S$ is the Hilbert space of the two-mode optical field and $\Q^1$ gives quadrature operators of the modes. For example, the modes can be Hermite-Gaussian modes of laser beam \cite{Siegman}. Or the modes can be spatially separated,  $\fii_0$ can be a coherent or squeezed state, and $\ket+$ could describe a laser field in the first mode when the second mode is not in use.

Finally, we note that a complete measurement is not necessarily informationally complete (IC) but if a POVM $\Mo$ is IC then $\Mo^1$ related to a complete measurement of $\Mo$ is also IC since one gets more information about the state of the system.
Recall that $\Mo$ is IC if the measurement statistics $p_\varrho$ determines the state $\varrho$, i.e.\ (in the discrete case) from 
$$
\tr{\varrho\Mo_{i}}\equiv\tr{\varrho'\Mo_{i}}
$$
follows that $\varrho=\varrho'$.
From Eq.\ \eqref{kljsdhklfh} we see that $\Mo_i=\sum_{k=1}^{m_i}\Mo^1_{ik}$ and thus $\tr{\varrho\Mo^1_{ik}}\equiv\tr{\varrho'\Mo^1_{ik}}$ implies $\tr{\varrho\Mo_{i}}\equiv\tr{\varrho'\Mo_{i}}$, i.e.\ if $\Mo$ is IC then $\Mo^1$ is also IC.

\vspace{1cm}

In conclusion, we have studied the role of the environment in quantum measurement theory considering a tripartite system $E+S+A$ where only the system $S$ and the ancilla $A$ interact during a local measurement of $S$, i.e.\ $S$ and $A$ are isolated from their environment $E$. 
If the (reduced) state of $S+A$ is not entangled, it does not mean that the state of the total system $E+S+A$ would be also separable.
If the measurement is complete then the post measurement state of the total system $E+S+A$ is automatically non-entangled.
Thus, complete measurements can be used to completely separate the system from its environment and to prepare non-entangled states. 
This has consequences e.g.\ to the quantum Zeno effect \cite{MiSu}:

Consider a quantum system $S$ (e.g.\  an atom) with the (discrete) Hamiltonian $H$ whose eigenstates are $d_{ik}$, i.e.\ $Hd_{ik}=x_id_{ik}$, $k=1,\ldots,m_i$, and $\<d_{ik}|d_{j\ell}\>=\delta_{ij}\delta_{k\ell}$. 
In practice, $S$ can never be totally isolated and will inevitably be an open system 
interacting with its environment $E$. 
If $E+S$ is prepared in a (entangled) pure state $\Psi$ and $x_i$ is obtained in the (von Neumann-L\"uders) measurement of $H$ then the (non normalized) output state is $\Psi_i=(I_E\otimes\Mo_i)\Psi$ where $\Mo_i=\sum_{k=1}^{m_i}\kb{d_{ik}}{d_{ik}}$.

Suppose that $E+S$ evolves according to a unitary time evolution $U_{ES}(t)$ 
caused e.g.\ by quantum decoherence,
and $U_{ES}(t)$ commutes with the projection $I_E\otimes\Mo_i$ with $m_i>1$.
If $S$ is observed `continuously' then the (possibly entangled) state at time $t$ is
$
U_{ES}(t)\Psi_i/\|\Psi_i\|
$
so that the evolution is not freezed. Only if we measure $H$ completely and frequently enough we can say with certainty that $S$ stays in the state $d_{ik}$ (if $(x_i,k)$ is obtained), the time evolution is suppressed, and the state of $E+S$ is factorized, i.e.\ $S$ is totally decoupled from its decohering environment.
In this case, we have obtained a {\it complete quantum Zeno effect}
which seems to deserve further systematic study.

\vspace{1cm}

%%%%%%%%%%%%%%%

\end{document}